\documentclass[%
aip,
amsmath,amssymb,
reprint,%
]{revtex4-1}

\usepackage{graphicx}% Include figure files
\usepackage{dcolumn}% Align table columns on decimal point
\usepackage{bm}% bold math
\usepackage[utf8]{inputenc}
\usepackage[T1]{fontenc}
\usepackage{mathptmx}
\usepackage{etoolbox}
\usepackage{xcolor}
\usepackage{amssymb,amsfonts,amsmath}
\usepackage{float}
\usepackage{epstopdf}
\usepackage{xcolor}
\usepackage{textcmds}
\usepackage{textcomp}
\usepackage{soul}% \st
\usepackage{csquotes}
\usepackage{amsbsy}
\usepackage{mathrsfs} % script-like, curvy letters.
\usepackage{bigints} % Big Integration
\usepackage{amsthm} % Theorem Proof
\usepackage[english]{babel}
\usepackage[shortlabels]{enumitem}
\usepackage{mathrsfs,bigints,mathtools,dsfont}
\usepackage[colorlinks=true,linkcolor=blue,citecolor=blue]{hyperref}%
\usepackage[toc]{appendix}

\makeatletter
\def\@email#1#2{%
	\endgroup
	\patchcmd{\titleblock@produce}
	{\frontmatter@RRAPformat}
	{\frontmatter@RRAPformat{\produce@RRAP{*#1\href{mailto:#2}{#2}}}\frontmatter@RRAPformat}
	{}{}
}%
\makeatother
\begin{document}

\title{Directional synchrony among self-propelled particles under spatial influence}
	\author{Suvam Pal}
\affiliation{Physics and Applied Mathematics Unit, Indian Statistical Institute, 203 B. T. Road, Kolkata 700108, India}

\author{Gourab Kumar Sar}
\affiliation{Physics and Applied Mathematics Unit, Indian Statistical Institute, 203 B. T. Road, Kolkata 700108, India}

\author{Dibakar Ghosh}
\affiliation{Physics and Applied Mathematics Unit, Indian Statistical Institute, 203 B. T. Road, Kolkata 700108, India} 
\thanks{Corresponding Auhtor: Dibakar Ghosh}
\email{diba.ghosh@gmail.com (D. Ghosh)}

\author{Arnab Pal}
\affiliation{The Institute of Mathematical Sciences, CIT Campus, Taramani,
Chennai 600113, India \& Homi Bhabha National Institute,
Training School Complex, Anushakti Nagar, Mumbai 400094, India}

	%\subject{epidemic spreading, complex network}
	
	%\keywords{Complex network, epidemic spreading, final outbreak size, test-kit}
	%\thanks{These two authors contributed equally}

	%%%% Abstract text to be placed here %%%%%%%%%%%%
	\begin{abstract}
		\par {Synchronization is one of the emerging collective phenomena in interacting particle systems. Its ubiquitous presence in nature, science, and technology has fascinated the scientific community over the decades. Moreover, a great deal of research has been, and
is still being, devoted to understand various physical aspects of the subject. In particular, the study of interacting \textit{active} particles has
led to exotic phase transitions in such systems which have opened up a new research front-line. Motivated by this line of work, in this paper, we study the directional synchrony among self-propelled particles. These particles move inside a bounded region, and crucially their directions are also coupled with spatial degrees of freedom. We assume that the directional coupling between two particles is influenced by the relative spatial distance which changes over time. Furthermore, the nature of the influence is considered to be both short and long-ranged. We explore the phase transition scenario in both the cases and propose an approximation technique which enables us to analytically find the critical transition point. The results are further supported with numerical simulations. Our results have potential importance in the study of active systems like bird flocks, fish schools and swarming robots where spatial influence plays a pertinent role.}
\end{abstract}
	
% \pacs {05.45.Xt, 05.45.Gg, 85.25.Cp, 87.19.lm}
\maketitle
	
\begin{quotation}
    Swarming and alignment are two rich collective behaviors in statistical physics and nonlinear dynamics. Understanding these behaviors has been a challenge to the researchers over the years.  These phenomena primarily depend on the interactions among the agents or potential landscape (attractive/repulsive force). In Vicsek\cite{Vicsek} model, the vision range of each mobile agent plays a crucial role to control the alignment among them in the presence of noise. In this paper, we have replaced the vision range of the particles by distance dependent coupling functions. This type of influence of spatial positions on the directions of the particles is generic and allows us to analytically calculate the critical transition point from a desychronized configuration to a synchronized one by performing mean-field approximation.
\end{quotation}
%\begin{fmtext}
%%%%%%%%%%%%%%%%%%%%%%%%%%%%%%%%%%%%%%%%%%%%%%%%%%%%%%
%%%%%%%%%%%%%%%%%%%%%%%%%%%%%%%%%%%%%
\section{Introduction}
The study of collective dynamics comes into picture when one tries to comprehend the behavior of interacting entities that stay or move together. Phenomena like {\it swarming} \cite{carrillo2010particle,okubo1986dynamical} and {\it flocking} \cite{reynolds1987flocks,toner1998flocks} are some of the most widespread ramifications of collective dynamics with multi-agent systems. A myriad of applications can be found in many biological systems like bacterial colonies \cite{czirok1996formation,kearns2010field}, bird flocks \cite{cavagna2010scale,hemelrijk2012schools}, shoaling behavior in fishes \cite{huth1994simulation,partridge1982structure} where the length scales span from micro (microns) to macro (kilometers). Specifically, it is known that local interaction among large number of individuals drives them to a state of {\it collective motion}, which is commonly known as swarming. Generally, these systems have been demonstrated using self-propelled particles (SPPs) where the particles consume energy from the surrounding environment to drive themselves. They move with nonzero self propulsion velocity and orientation. The situation where the velocity and direction of propagation are synchronized among the agents, is regarded as flocking in literature.

\par Several multi-agent systems in nature can show synchronization along with collective spatial dynamics. Simply put, synchronization can be perceived as a state of unison. In 1967, Winfree \cite{winfree1967biological} proposed a limit cycle oscillator model that shows critically synchronized behavior. Later, Kuramoto proposed a general model which was analytically tractable \cite{kuramoto1975self}. Kuramoto's seminal paper on synchronization opened up a new avenue of research that lead to further studies of this model with different network topologies \cite{rodrigues2016kuramoto,dorfler2011critical}, distributed frequencies \cite{sakaguchi1986soluble}, time-varying interactions \cite{ghosh2022synchronized}, attractive-repulsive couplings \cite{hong2011kuramoto}, multi-layer structures \cite{gambuzza2015intra}, higher-order interactions \cite{millan2020explosive} etc. Synchronization phenomena have also been rigorously studied with various amplitude oscillator models \cite{pikovsky2001universal,pecora1990synchronization,boccaletti2002synchronization,anwar2023synchronization} other than the regular Kuramoto phase oscillator. However, in most of these studies, the spatial positions of particles played a minimal role in the sense that they did not have an effect on the dynamics of synchronization. This gap was curtailed to some extent once researchers started studying {\it mobile oscillators} \cite{frasca2008synchronization,fujiwara2011synchronization,chowdhury2019synchronization,majhi2019emergence}. In there, the particles, which are more often termed as oscillators, move in space and interact among them depending on their spatial locations. More recently, extensive works have been done on model systems where the above-mentioned unidirectional influence of spatial positions on internal dynamics has been replaced by a bidirectional mutual feedback of spatial dynamics and internal dynamics on each other. These model systems, termed as {\it swarmalators} \cite{o2017oscillators,sar2022dynamics,o2019review,sar2023pinning,o2022collective,sar2022swarmalators,sar2023swarmalators}, are still very much in their infancy and require extensive study.

\par In this work, we are interested in the collective dynamics of SPPs with a sharp distinction that the spatial positions influence the direction of propagation. Similar study in this direction was first carried out by Vicsek et al. \cite{Vicsek} while studying phase transition among interacting active particles. The agents switched from a gas-like incoherent state to one in which the agents moved as a coherent flock beyond a critical density. It was assumed that there is a vision range for each particle and the direction of that particle is the mean direction of all the particles lying inside that vision range. At later times, many studies adopted this approach and considered short-range interaction of particles~\cite{peruani2010mobility,peruani2017hydrodynamic,grossmann2016superdiffusion}. However, the coupling strengths inside this interaction region were not affected by the spatial distance between the particles. We replace this type of implicit dependence by incorporating \textit{explicit functional dependence} of spatial positions to the direction of propagation or the orientation of the particles.

\par To interconnect the spatial movement and the orientation of the particles, we consider SPPs which are moving on a plane and introduce a spatial dependent Kuramoto-like phase equation. The orientation dynamics is influenced by the distance between the particles in two ways. In one case, the distance between two particles enhances their directional coupling and the effect is inverse in the other case. Here, the directional coupling essentially indicates the coupling among particles' direction. Due to the noise present in the system, particles move in randomized directions throughout the region for low coupling strength, but for sufficiently large coupling their orientations tend to align. We study this transition from a randomized state to an ordered state with a mix and match of theoretical and/or numerical analysis. In the thermodynamic limit, we analytically derive the effective coupling strength and critical noise amplitude for the synchronized state. We also study the rate of change of the order parameter near the critical noise strength.

\par The rest of the paper is arranged as follows: In section Sec.~\ref{sec2}, we introduce and describe the model for SPPs. We proceed with the theoretical analysis of our model in Sec.~\ref{sec3}, where we elaborately discuss the nature of spatial influence and its dependence on the exponent $\alpha$. In Sec.~\ref{sec4}, we study the behavior of macroscopic order parameter near the critical transition point by performing finite size scaling analysis on the model. Finally, we summarize our paper with concluding remarks in Sec.~\ref{sec5}.

\section{Model set-up of SPP}
\label{sec2}
We consider $N$ number of identical self-propelled particles moving inside a square region with area $L\times L$ under periodic boundary conditions with speed $v_0$. The spatial dynamics is governed by
\begin{equation}\label{Eq.1}
    \dot{\vec{R}}_i=v_0\left(\cos \theta_i \hat{x}+\sin \theta_i \hat{y}\right),
\end{equation}
for $i = 1,2,\cdots,N$, where $\vec{R}_i \equiv (x_i,y_i)$ stands for the position of the $i$-th particle, and $\hat{x}$ and $\hat{y}$ are the unit vectors along $x$ and  $y$ axes, respectively. Here $\theta_i$'s represent the internal dynamics (in our case, the direction of propagation) of the particles. The coupled internal dynamics is given by the following equation
%which are connected through Kuramoto interaction,
\begin{equation}\label{Eq.2}
    \dot{\theta}_i=\omega_i+\dfrac{K}{N}\sum_{j=1,j\neq i}^{N}\left|\vec{R}_j-\vec{R}_i\right|^\alpha \sin\left(\theta_j-\theta_i\right)+\zeta_i,
\end{equation}
where $\omega_i$ signifies the angular frequency of the ith particle and which is identical for all the particles, i.e., $\omega_i = \omega$,  for $i = 1,2,\cdots, N$. By moving to a proper frame of reference, we can set $\omega=0$ \cite{hong2011conformists}. The sinusoidal coupling among the directional component of the particles is inspired by the Kuramoto coupling \cite{kuramoto1975self}. The coupling strength $K$ ($>0$) determines whether the directions behave asynchronously or synchronously \cite{acebron2005kuramoto}. The magnitude of $K$ is further influenced by the spatial positions of the particles through the term $\left|\vec{R}_j-\vec{R}_i\right|$, which is the Euclidean distance between the $i$-th and $j$-th particle. The exponent $\alpha$ in Eq.~\eqref{Eq.2} dictates the nature of influence that is exerted by the spatial degrees of the particles on the directional coupling. For a positive value of $\alpha$, the effective coupling strength between two particles is proportional to the distance, i.e., the coupling is high for far away particles compared to the ones which are nearby. The opposite scenario occurs for a negative $\alpha$. With $\alpha = 0$, we get back the stochastic Kuramoto dynamics without the effect of spatial positions on directions. In the subsequent sections, we will study these two cases ($\alpha = 0$ and $\alpha \ne 0$) in details. In addition, we have also assumed that the SPPs are interacting in some thermal background. This is considered by adding a thermal noise $\zeta(t)$ to the dynamics. The thermal noise can be understood as a random force being generated from the random collisions of the background particles. Here, we take this to be a uniform white noise with the following statistical properties
\begin{align}
    \langle \zeta_i(t)\rangle=0,~~~\langle \zeta_i(t)\zeta_j(t^\prime\rangle=2D\delta_{ij}\delta(t-t^\prime),
\end{align}
where $D$ is the noise strength. The white noise signal is generated from a uniform distribution $\left[-\eta/2,\eta/2\right]$ so that $2D = \eta^2/12$. Initially, the spatial positions of the particles are uniformly distributed inside the square region $L\times L$, and $\theta_i$'s are chosen randomly from $[0,2\pi)$.

\section{Results}
\label{sec3}
In this section, we will centrally focus on the analysis of directional synchrony among the particles. Here, the coupling topology is all-to-all, and the density of the particles does not play a factor in achieving synchrony, unlike the Vicsek model \cite{Vicsek} where particles move with a unit interaction radius. The main control parameters are the coupling strength $K$ and the noise amplitude $\eta$. The values of these two parameters have an opposing effect on the system: $K$ brings synchrony, whereas $\eta$ brings randomness in the directions. Moreover, the system dynamics is dependent on the value of $\alpha$. As mentioned earlier, the nature of influence of spatial dynamics on directional coupling varies depending on the value of $\alpha$ -- below, we list three such cases
\begin{enumerate}
    \item $\alpha = 0$: No influence of spatial dynamics on directional coupling. This is the conventional Kuramoto dynamics with noise  \cite{kuramoto1975self}.
    \item $\alpha>0$: Coupling strength is enhanced between faraway particles and reduced when they are within a unit distance. Attraction is dominant in long-range than in short-range.
    \item $\alpha<0$: On the directional degree, nearby particles are coupled with a greater strength than the distant ones. This means attraction is dominant in short-range than in long-range.
\end{enumerate}
We will discuss these individual cases in the following subsections. We start with the case $\alpha = 0$ which can be considered as simple Kuramoto network model without the impact of spatial positions and in presence of noise.

\subsection{Case I: $\alpha=0$ (no impact of spatial positions)}
In this case, the angular degrees of freedom follow the Kuramoto dynamics and the positions of the particles do not affect the coupling. Eq.~\eqref{Eq.2} boils down to
\begin{equation}\label{Eq.3}
    \dot{\theta}_i=\dfrac{K}{N}\sum_{j=1}^{N} \sin\left(\theta_j-\theta_i\right)+\zeta_i.
\end{equation}
To investigate the synchronization phenomena in the directionaility, we introduce the following complex quantity
\begin{equation}\label{Eq.4}
    r e^{\Tilde{i} \psi} = \dfrac{1}{N}\sum_{j=1}^{N}e^{\Tilde{i} \theta_j}, \hspace{10pt} \Tilde{i} = \sqrt{-1}
\end{equation}
where $r$ is the well-known order parameter (a measure of synchrony) of the Kuramoto model \cite{acebron2005kuramoto}. By definition, the value of $r$ always lie within $0$ and $1$. Clearly, $r\approx 1$ means that the directions are oriented towards the same, i.e., they are completely synchronized. On the other hand, $r<1$ indicates asynchronous behavior among them. Here, $\psi$ is the mean or average of $\theta_i$-s. In terms of the order parameter,  Eq.~\eqref{Eq.3} reads
\begin{equation}\label{Eq.5}
    \dot{\theta_i}=Kr\sin \left(\psi-\theta_i\right)+\zeta_i.
\end{equation}
Equation~\eqref{Eq.5} shows that a mean field $Kr$ acting on each oscillator $\theta_i$. Here, we assume that $r$ is time-independent, which allows us to find a self-consistent equation for $r$ by substituting the solution of Eq.~\eqref{Eq.5} into Eq.~\eqref{Eq.4}. It is known that, near the phase transition point $r\propto (K-K_c)^{1/2}$ and $\psi = \omega_0 t + \psi_0$ \cite{sakaguchi}, which means $\psi$ rotates at a uniform frequency $\omega_0$. By going into the rotating frame of reference with frequency $\omega_0$, the mean or average angle of propagation can be set to zero (i.e., $\psi$ = 0), without loss of generality. Then the modified form of the Eq.~\eqref{Eq.5} becomes
\begin{equation}\label{Eq.6}
    \dot{\theta_i}=-Kr\sin \theta_i+\zeta_i.
\end{equation}
Equation~\eqref{Eq.6} is the well-known Langevin equation in angular space with sinusoidal force \cite{evans2012introduction}. Setting $\rho(\theta,t)$ as the probability density of a particle having direction $\theta$ at time $t$, we can write the corresponding Fokker-Planck equation as
\begin{equation}\label{Eq.8}
    \dfrac{\partial }{\partial t}\rho(\theta,t)=-\dfrac{\partial}{\partial \theta}\left[\left(-Kr\sin\theta \right)\rho(\theta,t)\right]+D\dfrac{\partial^2}{\partial \theta^2}\rho(\theta,t),
\end{equation}
where note that $\rho$ is a $2\pi$ periodic function and it is normalized, i.e., $\int_{0}^{2\pi}\rho(\theta,t) d\theta=1$. In the long time, we expect the system to reach a stationary state defined as $\rho_{st}(\theta)$ which satisfies the following relation
\begin{equation}\label{Eq.10}
    \dfrac{d}{d \theta}\left[\left(Kr\sin\theta \right)\rho_{st}(\theta)\right]+D\dfrac{d^2}{d \theta^2}\rho_{st}(\theta)=0.
\end{equation}
Since there is no loss of probability, Eq. (\ref{Eq.8}) can also be understood as a continuity equation $\frac{\partial \rho}{\partial t}+\frac{\partial J}{\partial \theta}=0$, where $J$ is the probability current or the flux. Within this description, in the stationary state, the  current takes the following form
\begin{equation}\label{Eq.12}
    J(\theta)=Kr\rho_{st}\sin \theta + D \dfrac{d}{d\theta}\rho_{st}.
\end{equation}
Here the probability density function in stationary state $\rho_{st}$ and the probability density current $J$ both are periodic in nature. Imposing periodic boundary condition on $\rho_{st}$ and $J$, we can write $\rho_{st}(\theta)=\rho_{st}(\theta + 2\pi)$ and $J(\theta)=J(\theta + 2\pi)$. To solve Eq.~\eqref{Eq.10} with these boundary conditions, we consider the following function,
\begin{equation}\label{Eq.15}
    f(\theta)=\exp\left[-\int_{0}^{\theta}\dfrac{Kr\sin\phi}{D}d\phi\right].
\end{equation}
Skipping details (see Appendix \ref{calculations}), we can write the probability density function for stationary state in the following form
\begin{equation}\label{Eq.17}
    \rho_{st}(\theta)=\dfrac{\exp\left(\dfrac{Kr}{D}\cos\theta\right)}{2\pi I_0\left(\dfrac{Kr}{D}\right)},
\end{equation}
where $I_0$ is Bessel function of first kind of zeroth order. 
In the $N\to \infty$ limit, we can compute the order parameter from the stationary probability density function
\begin{equation}\label{Eq.7}
    r=\lim_{t \to \infty}r(t)=\int_{0}^{2\pi}\exp(\Tilde{i}\theta)\rho_{st}(\theta) d\theta,
\end{equation}
where one just needs to care about the real part of the integral \cite{sakaguchi}. Performing the integral one finds
\begin{align}
r=
I_1\left(\dfrac{Kr}{D}\right)\bigg/I_0\left(\dfrac{Kr}{D}\right)\label{Eq.19},
\end{align}
where $I_1$ is Bessel function of first kind of first order.
Near the phase transition point, i.e., $r\rightarrow 0$ limit, $I_0$ and $I_1$ can be expanded in series with respect to 
$K_cr/D$, where $K_c$ is the critical coupling strength that marks the continuous transition from sync state to async state. Neglecting higher order terms, Eq.~\eqref{Eq.19} can be reduced to the following form
\begin{equation}\label{Eq.20}
    K_c=2D.
\end{equation}
The above analysis provides a distinct critical coupling strength for $\alpha=0$. Drawing insights from this, we would like to explore whether a similar analysis can be done for the $\alpha\ne 0$ case. However, the spatial coupling on the orientations renders a new layer of complexity. Thus, moving forward, we apply rational approximations to the overall dynamics of the system which allows us to make some coherent inference. Next, we discuss these in details.

\subsection{Case II: $\alpha\neq 0$ (under the influence of spatial positions)}\label{subsec,a}
The self-propelled particles roam inside the bounded region following the dynamics \eqref{Eq.1}. The spatial coordinates do not interact among them. Thus, at a given time, the spatial coordinates inside the region remain uniformly distributed. This might not be the case with nonzero and nonidentical frequency distributions of the particles as one can expect chiral pattern for large frequency values~\cite{liebchen2017collective,liebchen2022chiral}. Since the frequency of the particles is set to zero in our model, we do not encounter this issue and numerics also confirm that the spatial distribution remains to be uniform. To avoid the occurrence of collision among them, we fix a lower bound on the inter-particle distance, say $\Delta(\alpha)$. This ensures that at any moment they can not come closer than $\Delta(\alpha)$ and thus collision among them is safely disregarded. We employ the following condition along with Eq.~\eqref{Eq.1}: Suppose at time $t$ the spatial distance between the $i$-th and $j$-th particles is $d_{ij}(t)$. Then, we take $d_{ij}(t) = \max \{d_{ij}(t), \Delta(\alpha) \}$. It should be noted that the spatial dynamics of Eq.~\eqref{Eq.1} does not contain any attraction or repulsion forces among the particles' positions unlike other swarming models or cell dynamical models \cite{fetecau2011swarm}. Since the particles move inside a bounded domain, the maximum distance between any two particles inside a square box is the diagonal of the square region. But under the periodic boundary condition and minimum image convention, all four vertices of the region are identical. Thus, the maximum distance between two particles essentially reduces to half of the diagonal length, which is $L/\sqrt{2}$.

\par To seek the condition for which the particles are synchronized, we rewrite Eq.~\eqref{Eq.2} as
\begin{equation}\label{Eq.21}
    \dot{\theta}_i = \dfrac{1}{N}\sum_{j=1,j\neq i}^{N}K_{ij} \sin\left(\theta_j-\theta_i\right)+\zeta_i,
\end{equation}
where $K_{ij} = K \left|\vec{R}_j-\vec{R}_i\right|^{\alpha}$ can be considered as time-varying weighted coupling strength between the $i$-th and the $j$-th particle. Since the distances are uniformly distributed in the bounded region $L \times L$, so are the $K_{ij}$'s. We can next take the annealed approximation such that $\sum_{j\neq i}K_{ij} = \langle K_{ij} \rangle \sum_{j\neq i}$, where $\langle K_{ij} \rangle$ is the mean of the distributed coupling strengths. We consider $\langle K_{ij} \rangle$ as the effective coupling strength and denote it as $K^{eff}(\alpha)$ and it depends on the choice of $\alpha$. Equation~\eqref{Eq.21} in terms of effective coupling strength becomes,
\begin{equation}\label{Eq.22}
    \dot{\theta}_i\approx\dfrac{K^{eff}(\alpha)}{N}\sum_{j=1,j\neq i}^{N} \sin\left(\theta_j-\theta_i\right)+\zeta_i.
\end{equation}
In the continuum limit $N\rightarrow \infty$, $K^{eff}(\alpha)$ can be calculated in terms of a finite integral
\begin{equation}\label{Eq.23}
    K^{eff}(\alpha)= \dfrac{K}{L/\sqrt{2}-\Delta(\alpha)}\int_{\Delta(\alpha)}^{L/\sqrt{2}}x^{\alpha} dx.
\end{equation}
Now, depending on the value of $\alpha$, the effective coupling among particles varies. It also has a dependence on the box size $L$. Next, we investigate the phase transition behavior for the cases $\alpha>0$ and $\alpha<0$, one by one.

\par Before delving deeper into the analysis, we can qualitatively gain some insights of these two cases. Since we are considering a bounded squared region of length $L$, the area under consideration is $L^2$. For a given particle, the strength of coupling with others depends crucially on their distances from that particle, in particular, to the fact that whether they lie within a unit distance or outside. The ratio of number of particles lying inside to the same lying outside is $\dfrac{\pi}{L^2 - \pi}$, which is less than 1 for $L> \sqrt{2 \pi}$. For $\alpha>0$, particles which are lying outside a unit distance are coupled with higher strength than those lying inside. This essentially means that more number of particles are coupled with higher coupling strengths. A converse scenario appears for $\alpha<0$. In here, the particles lying inside a unit distance are coupled with higher strengths. So, a comparatively lesser number of particles are coupled with larger strengths. We infer that the particles can attain synchrony faster with $\alpha>0$ than with $\alpha<0$ as one changes the coupling strength $K$. Keeping these insights in mind, we now proceed to study these cases both analytically and numerically.

\subsubsection{$\alpha>0$ : Long-range dominance of coupling}
In this case, the phase coupling strength between two particles is enhanced when their distance is greater than unity and is reduced otherwise. In other words, the attraction between the particles is more effective in the long-distance than in the short-distance. Here, we can safely allow the particles to come asymptotically close to each other without colliding, i.e., $\Delta(\alpha)\rightarrow 0$ since the interaction term appears only in the numerator. Within this assumption, the integral on the right hand side of Eq.~\eqref{Eq.23} can be calculated easily and the effective coupling strength becomes
\begin{equation}\label{Eq.24}
    K^{eff}(\alpha)= \dfrac{K}{\alpha+1} \Big(\dfrac{L}{\sqrt{2}}\Big)^{\alpha}.
\end{equation}
This phase coupling strength $K^{eff}(\alpha)$ among the particles is independent of the inter-particle distances. Then the model reduces to one where the particles interact among themselves with a constant coupling like in the case of $\alpha=0$ in Eq.~\eqref{Eq.3}. It enables us to write down the critical coupling strength using Eq.~\eqref{Eq.20} as
\begin{equation}\label{Eq.25}
    K^{eff}_c(\alpha) = 2D,
\end{equation}
where $2D={\eta}^2/12$. Substituting Eq.~\eqref{Eq.24} into Eq.~\eqref{Eq.25}, we further get
\begin{equation}\label{Eq.26}
    \dfrac{K_c}{\alpha+1} \Big(\dfrac{L}{\sqrt{2}}\Big)^{\alpha} = \dfrac{\eta_c^2}{12},
\end{equation}
where $K_c$ and $\eta_c$ are the critical coupling strength and critical noise amplitude, respectively. Next essential step is to check the validation of this mean field analysis comparing with simulations. To see this, we set $\alpha=1$ in Eq.~\eqref{Eq.26} to get
\begin{equation}\label{Eq.27}
    \eta_c^2=3\sqrt{2}K_cL.
\end{equation}
The presence of the component $L$ in the Eq.~\eqref{Eq.27} clearly indicates the effect of box size on the synchronization condition. We validate this fact by plotting the order parameter $r$ as a function of $\eta$ for two different values of $L$ with a fixed $K=0.05$. In Fig.~\ref{pt,a=1}, the blue and red colors demonstrate the results for $L=10$ and $L=20$, respectively. The solid lines are the $r$ vs. $\eta$ curves calculated numerically from the exact equations without approximation. The blue stars and red dots are those calculated from the model after performing approximation. When we increase $\eta$, randomness in the directionality grows, which effectively leads to a disordered state. It is thus evident that there is a transition from the ordered state (i.e., synchronization) to the disordered (asynchronous) state. We refer to Supplementary Movies 1 \& 2 which illustrate the evolution of particles in the desynchronized and synchronized states, respectively. Subsequently, the value of order parameter $r$ is monotonically decreasing with increasing $\eta$. We find the critical $\eta$ values from Eq.~\eqref{Eq.27} for $L=10$ and $L=20$ as $\eta_c \approx 1.47$ and $\eta_c \approx 2.06$, respectively. These match pretty well with our numerical curves. The small fluctuations are due to the finite particle number since the approximation is performed in the $N \to \infty$ limit. Fig.~\ref{pt,a=1} highlights three facts: (i) the obvious dependence of the transition curve on box size $L$; (ii) the accuracy of our approximation; and (iii) the agreement of numerical and analytical results. 

\begin{figure}[t]
    \centering
    \includegraphics[width=\columnwidth]{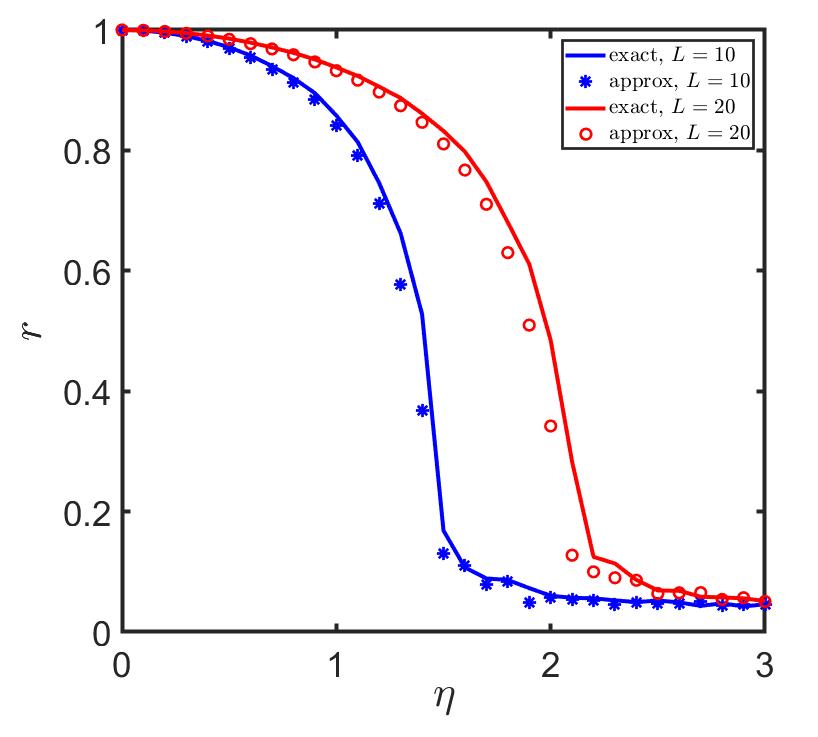}
    \caption{\textbf{Order parameter $r$ as a function of $\eta$ for different box sizes $L$ for $\alpha=1$.} Variation of $r$ has been shown as a function of $\eta$ for a fixed coupling strength $K=0.05$. We perform simulations with velocity $v_0=0.2$ and $N=500$ over $500$ time units with step size $0.1$ using Heun's method. The initial positions of the particles are uniformly distributed inside the domain and their directions are also chosen randomly from $[0,2\pi]$.  We calculate $r$ over last $5 \%$ data after discarding the transients and it is further averaged with five different initial configurations. Solid lines (in blue and red) represent numerically calculated values of $r$ from Eqs.~\eqref{Eq.1}-\eqref{Eq.2}, for $L=10$ and $20$, respectively, which we term as \textit{exact}. Blue stars and red circles are the values of $r$ calculated numerically from the mean-field analysis and we term them as \textit{approx}. We observe that the critical noise amplitude $\eta_c$ is enhanced as the box size increases. $\eta_c\approx1.47$ and $\eta_c\approx2.06$ for $L=10$ and $L=20$, respectively.}
    \label{pt,a=1}
\end{figure}

\par In Fig.~\ref{pt,a=1}, we check the transition phenomena with a fixed value of $K$. Having observed the effectiveness of our approximation and the agreement between numerics and analytical results, we intend to check if it is also valid for arbitrary coupling strength. For that, we move to a two-dimensional phase space (mesh) where we vary both $\eta$ and $K$ within some given intervals, and at each of the mesh points, we numerically calculate the value of the order parameter $r$. Then the points are colored corresponding to their respective $r$ values. We perform this simulation on the original model (i.e., without approximation) by taking $K$ from the interval $[0,0.2]$ and $\eta$ from $[0,3]$. We have done this by considering a $100\times 100$ mesh. The result is presented in Fig.~\ref{ps,alpha>0}(a). In the color bar, yellow indicates synchrony in directions, and blue indicates asynchrony. The synchronization and desynchronization regions separate themselves as they are colored accordingly. We draw the solid curve (in black) from Eq.~(\ref{Eq.27}) which stands for the critical point of transition. It is quite evident that the numerical result from the original model, Eq.~\eqref{Eq.2}, is well supported by the analytical findings on the approximated model, Eq.~\eqref{Eq.22}.

\begin{figure}[htp]
    \centering
    \includegraphics[width=\columnwidth]{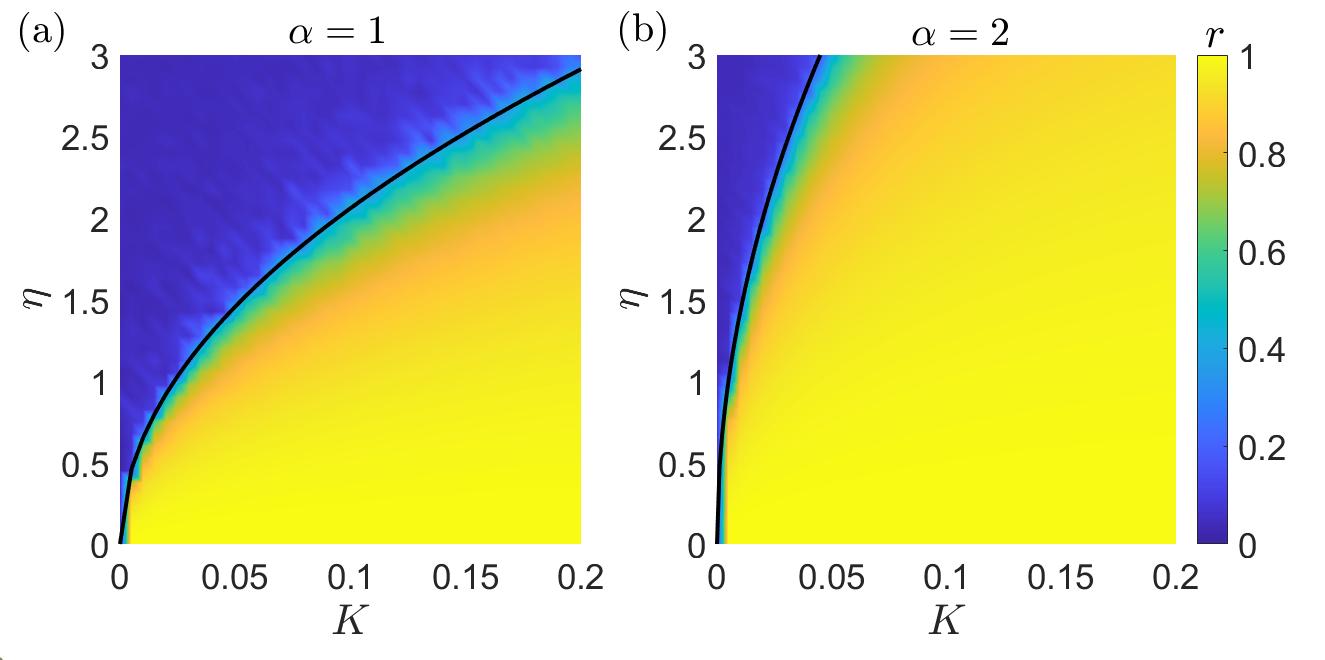}
    \caption{\textbf{Contour plot of order parameter $r$ in the $K$-$\eta$ parameter plane for $\alpha>0$.} Here, we choose $v_0 = 0.2$, $L=10$, and $N = 500$. We divide the $K$-$\eta$ plane in $100 \times 100$ mesh points and calculate the value of $r$ at each of these points. Data is generated by integrating the Eqs.~\eqref{Eq.1}-\eqref{Eq.2} using Heun's method with a step-size 0.1 for $t=2000$ time units. And order parameter $r$ is calculated by taking the average over the last $5 \%$ data and further averaging it over $5$ realizations. Left panel (a) $\alpha=1$: The solid curve (in black) is the analytical prediction drawn using Eq.~\eqref{Eq.27}. The region in yellow is the synchronized state whereas the blue region is the desynchronized one. Right panel (b) $\alpha=2$: The black curve is drawn using Eq.~\eqref{Eq.26} with $\alpha=2$.}
    \label{ps,alpha>0}
\end{figure}

\par To see the robustness of our analysis, we repeat our study for a different value of $\alpha >0$. For $\alpha=2$, the critical curve Eq.~\eqref{Eq.26} becomes $\eta_c^2 = 2 K_c L^2$. This is represented in Fig.~\ref{ps,alpha>0}(b) by the solid curve (in black). It acts as a separatrix between the synchronized and the desynchronized regions. Here too, we find an excellent argument between the numerical simulation and the mean-field analysis. 

\subsubsection{$\alpha<0$ : Short-range dominance of coupling}
Contrary to the previous case, for $\alpha<0$, the directional coupling strength between two particles is enhanced when they lie within a unit distance. On the other hand, if the distance between them is greater than unity, the coupling strength is reduced. The coupling strength keeps increasing as long as the distance between particles decreases. However, we can not allow the particles to come asymptotically close to each other
as the coupling becomes unfeasible. To avoid this scenario, we set a cut-off distance $\Delta(\alpha)$ for the particles. We specifically choose $\Delta(\alpha)=(0.1)^{1/|\alpha|}$ so that $\left|\vec{R}_j-\vec{R}_i\right|^{\alpha}$ always remains greater than $0.1$ making the system well-bounded without any ambiguity.

\par We again begin with the approximated expression for effective coupling strength as proposed in Eq.~\eqref{Eq.23} to study the directional synchronization. In this case, the effective coupling strength becomes
\begin{align}\label{Eq.28}
    K^{eff} = \begin{cases}
  \dfrac{K}{L/\sqrt{2}-\Delta(\alpha)} \ln\Big[\dfrac{L}{\sqrt{2}\Delta(\alpha)}\Big],  & \alpha = -1 \\
  \dfrac{K}{L/\sqrt{2}-\Delta(\alpha)} \dfrac{(L/\sqrt{2})^{\alpha+1} - \Delta^{\alpha+1}(\alpha)}{\alpha+1}, & \alpha \ne -1.
\end{cases}
\end{align}
For $\alpha=-1$, using Eq.~\eqref{Eq.28}, the critical value of noise amplitude becomes
\begin{equation}\label{Eq.29}
    \eta_c^2=\dfrac{12K_c}{L/\sqrt{2}-\Delta(-1)}\ln\Big[{\dfrac{L}{\sqrt{2}\Delta(-1)}}\Big].
\end{equation}
We verify this condition by performing numerical simulations. The set-up is similar as before. At each point of the $2$d mesh in the $K$-$\eta$ plane, the value of $r$ is calculated. The point is color coded according to it. The result is shown in Fig.~\ref{ps,alpha<0}(a). The points in yellow denote synchrony, whereas blue color coded points denote asynchrony. The most visible difference from the case $\alpha=1$ (as in Fig.~\ref{ps,alpha>0}(a)) appears as a significant reduction of the synchronized region. This is because the coupling strength is only dominant inside a unit distance. The solid curve (black) in Fig.~\ref{ps,alpha<0}(a) represents the analytical critical curve, Eq.~\eqref{Eq.29}. This is again fairly close to the transition region and strengthens the validity of our approximated mean field like method. We refer to Supplementary Movies 3 \& 4 which respectively show the time evolution of particles in the disordered and ordered states when $\alpha=-1$.

\par We also verify our approximation with $\alpha=-2$. After putting $\alpha=-2$ in the Eq.~\eqref{Eq.28} and taking $\Delta(\alpha=-2)=0.1^{1/2}$, Eq.~\eqref{Eq.28} reduces to
\begin{equation}\label{Eq.B1}
    K^{eff}=\dfrac{K}{L/\sqrt{2}-\sqrt{0.1}}\left(\dfrac{1}{\sqrt{0.1}}-\dfrac{\sqrt{2}}{L}\right).
\end{equation}
From Eqs.~\eqref{Eq.25} and \eqref{Eq.B1}, the expression for separatrix becomes,
\begin{equation}\label{Eq.B2}
    \eta_c^2=\dfrac{12K_c}{L/\sqrt{2}-\sqrt{0.1}}\left(\dfrac{1}{\sqrt{0.1}}-\dfrac{\sqrt{2}}{L}\right),
\end{equation}
which is again plotted by a solid curve (black) in Fig.~\ref{ps,alpha<0}(b).

\begin{figure}[t]
    \centering
    \includegraphics[width=\linewidth]{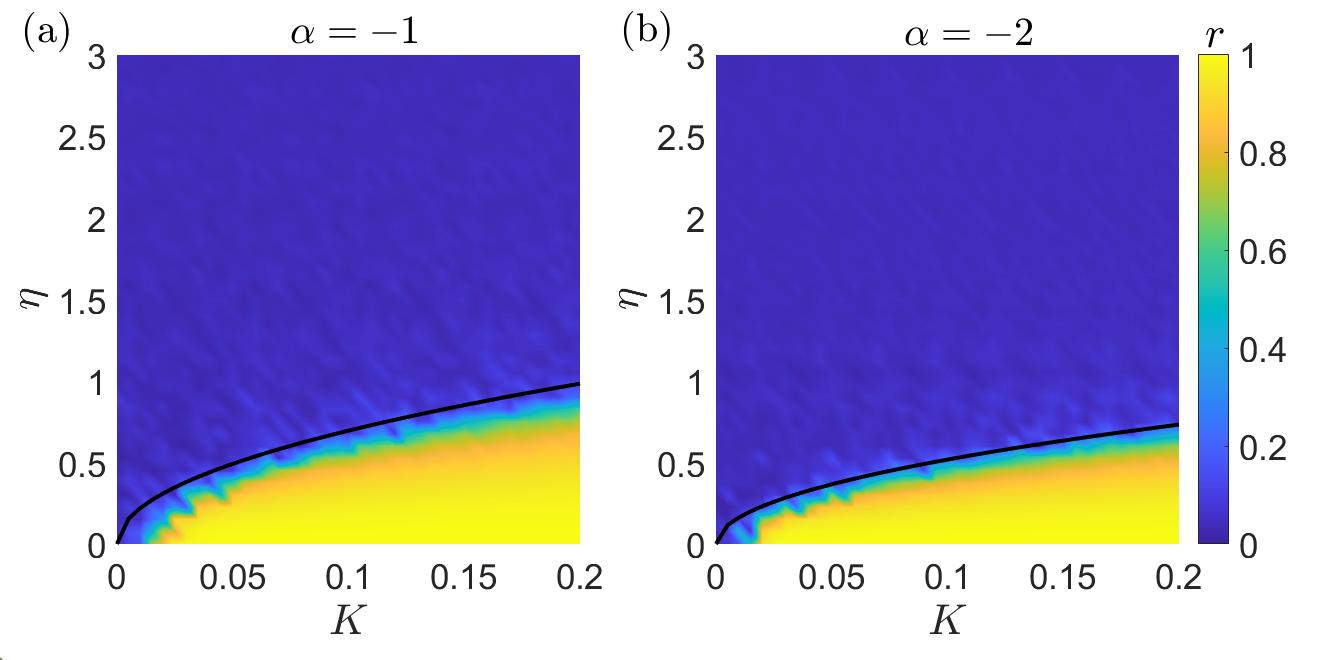}
    \caption{\textbf{Contour plot of order parameter $r$ in the $K$-$\eta$ parameter plane for $\alpha<0$.} We divide the $K$-$\eta$ plane in $100 \times 100$ mesh points and calculate the value of $r$ at each of these points. Data is generated by integrating the Eqs.~\eqref{Eq.1}-\eqref{Eq.2} using Heun's method with a step-size 0.1 for $t=2000$ time units. And order parameter $r$ is calculated by taking the average over the last $5 \%$ data and further averaging it over $5$ realizations. Bifurcation diagram for (a) $\alpha=-1$, and (b) $\alpha=-2$. Here, compared to $\alpha=1$ case in Fig.~\ref{ps,alpha>0}, the synchronized region (in yellow) is significantly reduced. The asynchronized region in blue prevails over most of the $K$-$\eta$ plane. The separatrices are represented with solid lines (in black) using Eqs.~\eqref{Eq.29} and \eqref{Eq.B2} respectively, in panel (a) and panel (b). Other parameters are $v_0=0.2$, $L=10$, and $N=500$.}
    \label{ps,alpha<0}
\end{figure}

\par So far we have established the numerical results and validated our approximations with them for both $\alpha>0$ and $\alpha<0$. In both cases, we find the critical transition points from the analytical expressions. The transition points depend on the system size $L$, we verify this numerically by varying the system size. In Fig.~\ref{scaling,all}, we draw critical coupling strength $K_c$ as function of $L$. The theoretical curves are indicated by the blue lines derived from Eq.~\eqref{Eq.29} (for $\alpha=-1$) and Eq.~\eqref{Eq.27} (for $\alpha=1$). Red circles are the numerically calculated critical coupling strengths using Binder's cumulant (see Sec.~\ref{sec4} for details). In the next section, we study the behavior of order parameter $r$ near the transition point by performing finite size scaling analysis.

\begin{figure}[t]
            \centering
            \includegraphics[width=\columnwidth]{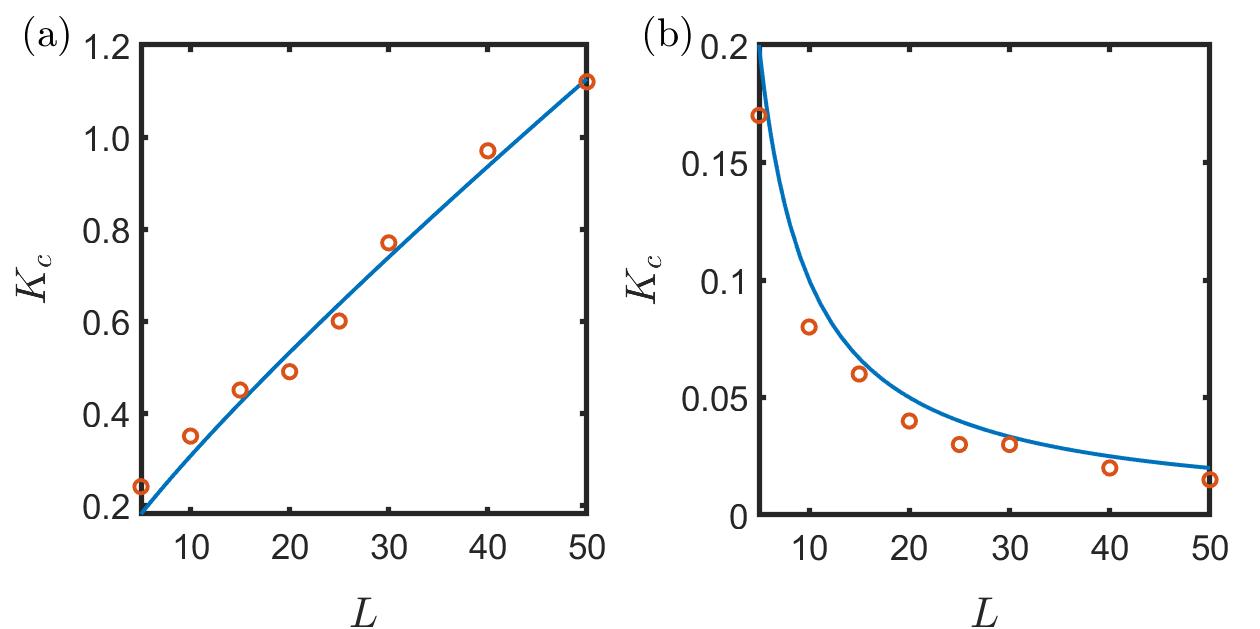}
            \caption{\textbf{Critical coupling strength $K_c$ as a function of the system size $L$.} The noise strength $\eta$ is kept fixed here. Solid lines (in blue) represent the theoretical values, whereas red circles represent the numerical values of $K_c$ evaluated numerically using Binder's cumulant. (a) The behavior of $K_c$ with respect to $L$ for $\alpha=-1$ with noise strength $\eta=1.5$. (b) Represents the behavior of $K_c$ with respect to $L$ for $\alpha=1$ with noise strength $\eta=2.06$. Initially, particles are placed uniformly randomly inside a box with size $L\times L$ and directions are chosen randomly from $[0,2\pi]$. Here, $(t,dt,N)=(500,0.1,1000)$. The macroscopic order parameter $r$ is calculated using last $5\%$ data and is averaged over $5$ realizations of initial conditions.}
            \label{scaling,all}
\end{figure}

\section{Finite size scaling analysis}
\label{sec4}
In thermodynamic limit, our system exhibits a continuous phase transition. Near the transition point, variation of the macroscopic order parameter becomes sensitive to the noise amplitude. To examine this, we make the following ansatz for the rate of variation
\begin{equation}\label{Eq.30}
    r \sim (\eta_c-\eta)^\beta,
\end{equation}
where $\beta$ signifies how fast $r$ varies near the critical point. We have already calculated the critical noise amplitude analytically in the previous section. Eqs.~\eqref{Eq.27} and \eqref{Eq.29} are the expressions of $\eta_c$ for $\alpha=1$ and -1, respectively. We can confirm this critical value by means of the well-known estimator \textit{Binder's cumulant} \cite{vollmayr1993finite} which detects the behavior of the phase transition. This is essentially the kurtosis of the macroscopic order parameter given by 
\begin{equation}
    U_4=1-\dfrac{\langle r^4\rangle}{3\langle r^2\rangle^2},
    \label{Eq.31}
\end{equation}
where $\langle \cdots \rangle$ signifies the ensemble average of initial configurations. For small fluctuations in $r$, we can approximate $\langle r^4\rangle \approx \langle r^2\rangle^2$ and thus, $U_4$ lies close to $\dfrac{2}{3}$. But near the transition point, where the fluctuation in $r$ is large, $U_4$ takes a jump. The value of the tuning parameter at which $U_4$ takes the minimum value, is usually considered to be the critical point \cite{vollmayr1993finite}. In first-order transition, the variation of macroscopic order parameter shows a discontinuous transition with respect to any external tuning parameter and $U_4$ takes a sharp jump to a very large negative value at the transition point. But in the second order phase transition, the variation of order parameter shows a continuous transition and $U_4$ takes a jump but always remains positive. We have calculated $U_4$ for $\alpha=1$ with $K=0.05$ while varying $\eta$. The results are accumulated in Fig.~\ref{bc,a=1}. Star indicators (in red) indicate the values of $U_4$ and the solid curve (in blue) stands for the order parameter $r$ numerically calculated from the model. We find that $U_4$ takes a jump around $\eta \approx 1.47$ where $r$ also goes through a visible change from a higher value to a lower value. The fact that $U_4$ remains positive near the transition indicates a second order phase transition. Using Eq.~\eqref{Eq.27}, we find $\eta_c=1.5$ for $\alpha=1$, $K=0.05$, and $L=10$. This is satisfactorily close to the one we achieved by performing the $U_4$ analysis. By doing a similar analysis for $\alpha=-1$, we find critical noise amplitude $\eta_c=0.54$ for coupling strength $K=0.05$.

\begin{figure}[t]
    \centering
    \includegraphics[width=\linewidth]{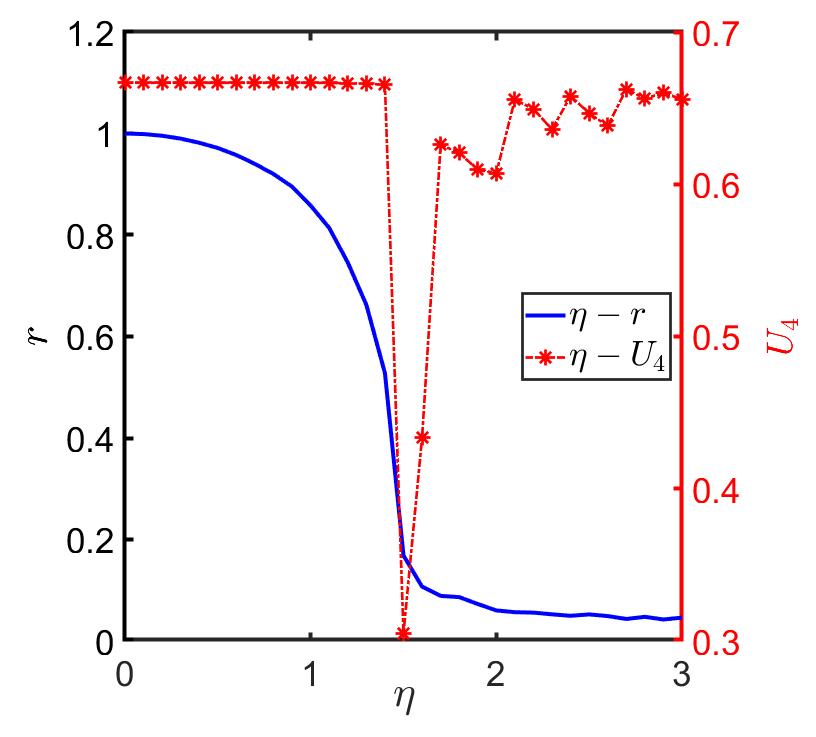}
    \caption{{\bf Finding the critical noise amplitude using Binder's cumulant.} We simulate Eqs.~\eqref{Eq.1} \& \eqref{Eq.2} numerically to calculate the value of $r$ at each $\eta$ using Heun's method. Data is generated with a step-size 0.1 for $t=2000$ time units. $r$ is calculated by taking the average over the last $5 \%$ data and further averaging it over $5$ realizations. Star indicators (red) are the values of $U_4$, given in Eq.~\eqref{Eq.31}. They are joined by dashed line to show the variation of $r$ with respect to $\eta$ for $\alpha=1$. The scale of variation is indicated in the right $y$ axis. The lower bound value of the jump remains positive which essentially confirms that the phase transition is of second order in nature. Lowest value occurs at $\eta_c\approx 1.47$ which closely matches with the analytical result, Eq.~\eqref{Eq.27}. Here, we have taken $K=0.05$ and $L=10$. Blue curve signifies the variation of $r$ inside the same range of $\eta$. $r$ varies between 0 to 1 which is shown in the left $y$ axis. Here, $v_0=0.2$ and $N=500$.}
    \label{bc,a=1}
\end{figure}

\par Finally, we try to find out the dependency of the noise amplitude on the macroscopic order parameter for fixed coupling strength near the phase transition. In Fig.~\ref{fs,a=1,a=-1}, the slopes of the curves indicate the values of $\beta$. As mentioned earlier, for a fixed coupling strength, the system with exponent $\alpha=1$ can sustain synchrony for larger value of noise amplitude than $\alpha=-1$. The curves in Fig.~\ref{fs,a=1,a=-1} show that the slope for the system with $\alpha=-1$ is larger than $\alpha=1$ thus conferring to the same argument.

\begin{figure}[t]
    \centering
    \includegraphics[width=\linewidth]{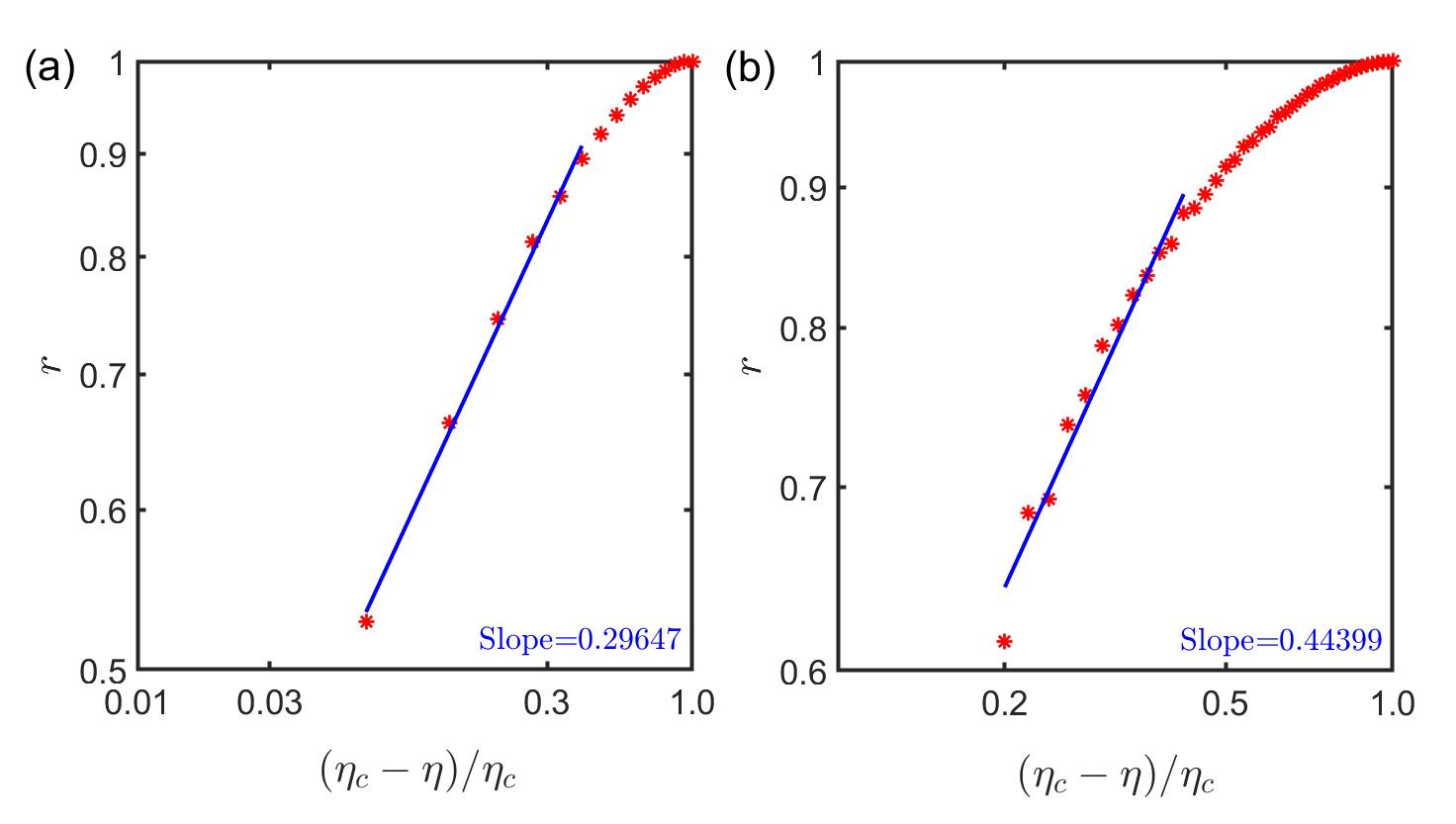}
    \caption{\textbf{Behavior of order parameter near the transition point}. The decay of the order parameter $r$  has been examined for different $\alpha$-s but for a fixed coupling strength. The particles are moving inside a bounded region with box length $10$ units and the coupling strength among them is fixed at $K=0.05$. Using the description of Binder's cumulant, we estimate the values of critical noise amplitude $\eta_c$. Panels (a) and (b) respectively describe the phenomena for $\alpha=1$ and $-1$ and the critical noise amplitudes are found to be $\eta_c\approx 1.47$ and $0.54$. By finite size scaling, we get $\beta\approx 0.29$ for $\alpha=1$ and $\beta\approx 0.44$ for $\alpha=-1$. For simulations, we use $v_0=0.2$ and $N=500$.}
    \label{fs,a=1,a=-1}
\end{figure}

\section{Conclusions}
\label{sec5}
In summary, we have studied the directional synchrony of SPPs inside a confined region under periodic boundary conditions. The key input here is the introduction of the spatial dependence among the particles in the orientational degrees of freedom and to demonstrate how they played a crucial role in effective coupling and in the phase transition. First we benchmarked our model in the absence of spatial influence and found the critical condition for synchronization to be $K_c = 2D$. Under the impact of spatial influence, we took annealed approximation of the distributed coupling strengths which enabled us in calculating the effective coupling strength. We also showcased our numerical results which are in excellent agreement with the analytical findings. According to our simulations,
in a wide range of the velocities of order from $10^{-3}$ to $10^{-1}$, the value of $v_0$ does not alter the results. %For our results, we used $v_0=0.2$.

\par In this detailed study, we considered both the cases where coupling between two particles is strong when they are nearby and the other way around. In one case, the spatial influence enhances the synchronization (for $\alpha>0$) and an opposite effect is observed otherwise (for $\alpha<0$). Earlier studies have considered the effect of mobility of the particles on achieving synchronization when they move randomly on one-dimensional and two-dimensional lattices~\cite{uriu2010random}. Over the years, a persistent effort has been made to understand the effect of spatial positions and mobility on the synchronization behavior and our work is one such attempt towards this goal. Intuitively, our model can mimic the properties of Vicsek model \cite{Vicsek} for a reasonable negative value of $\alpha$. Only the particles inside a unit vision radius are shown to be coupled with significant coupling strength, and outside it the coupling strength becomes negligibly small. In a nutshell, our model captures the essential characteristics of active matter systems that follow Vicsek type interactions.

\par Our model, \textit{albeit} general, has certain limitations that open up the stage for future studies. For example, we assume that the particles can collide among themselves unless some external condition is imposed on them. In our work, this drawback was taken care of by invoking a minimal distance $\Delta(\alpha)$ among the particles. However, collision exclusion is a necessary condition for the study of flocking and many other natural set-ups. Thus, attempts should be made by considering intrinsic spatial attraction and repulsion among the particles with this inherent condition. It is convenient to mention here that, although our mean-field approximation gives fairly satisfactory result, there can be other methods for calculating the phase transition curves. One such method can be to consider the local order parameter depending on the spatial position and incorporate it in the Fokker-Planck equation \cite{peruani2010mobility}. A field theory approach using tools from statistical mechanics can also be useful~\cite{grossmann2016superdiffusion}. In here, we have assumed zero frequency, but many interesting questions arise when one also considers nonzero angular frequencies in these models namely the chiral active matter \cite{liebchen2017collective,liebchen2022chiral}. Another interesting direction would be to consider other physically amenable forms for the spatial interaction. The analytical tractability of our work provides a testing ground for attempting these questions in near future. 

%%%%%%%%%%%

\section*{Acknowledgment}
Arnab Pal gratefully acknowledges research support from the Department of Science and Technology, India, SERB Start-up Research Grant Number SRG/2022/000080.

\section*{\label{sec:level7}Data Availability}
Data sharing is not applicable to this article as no new data were created or analyzed in this study.

\appendix
\section{Derivation of the steady state in Eq.~\eqref{Eq.17}}\label{calculations}
In this section we sketch out the steps to derive Eq.~\eqref{Eq.17}. To this end, let us recall the set-up where we have assumed that the orientation of a particle diffuses in the presence of a nonlinear force. The Fokker-Planck equation for such process reads
\begin{equation}\label{1a}
    \dfrac{\partial \rho(\theta,t)}{\partial t}=-\dfrac{\partial}{\partial \theta}\left[A(\theta)\rho(\theta,t)\right]+\dfrac{1}{2}\dfrac{\partial^2}{\partial \theta^2}\left[B(\theta)\rho(\theta,t)\right].  
\end{equation}
In the steady state, setting $\dfrac{\partial \rho(\theta,t)}{\partial t}=0$, we find
\begin{equation}\label{2a}
    -\dfrac{d}{d\theta}\left[A(\theta)\rho_{st}(\theta)\right]+\dfrac{1}{2}\dfrac{d^2}{d\theta^2}\left[B(\theta)\rho_{st}(\theta)\right]=0,
\end{equation}
where $\rho_{st}(\theta)$ is the time independent stationary state. From the continuity equation, one can easily identify the probability current in the steady state which reads
\begin{equation}\label{3a}
    J(\theta)=A(\theta)\rho_{st}(\theta)-\dfrac{1}{2}\dfrac{d}{d\theta}\left[B(\theta)\rho_{st}(\theta)\right].
\end{equation}
Both the stationary state and the currents are periodic in nature so that
\begin{align}
    \rho_{st}(\theta)=\rho_{st}(\theta+2\pi),~~~~J(\theta)=J(\theta+2\pi).
\end{align}
In the stationary state, $\partial J/\partial \theta=0$ which implies $J$ is constant throughout the space. To deal with the boundary conditions, we consider the following function
\begin{equation}\label{4a}
    f(\theta)=\exp\left[2\int_{0}^{\theta}\dfrac{A(\phi)}{B(\phi)}d\phi\right].
\end{equation}
A first differentiation with respect to $\theta$ yields
\begin{equation}\label{5a}
    f^\prime(\theta)=2 \dfrac{A(\theta)}{B(\theta)}f(\theta).
\end{equation}
Let us now also consider the following relation
\begin{equation}\label{6a}
    \dfrac{d}{d\theta}\left[\dfrac{B(\theta)\rho_{st}(\theta)}{f(\theta)}\right]=-\dfrac{2J(\theta)}{f(\theta)},
\end{equation}
where we have used the explicit form of the current from Eq. (\ref{3a}) and also used the relation above. Integrating both sides of the above equation and using the fact that $J(\theta)=J$ in the stationary state, we find
\begin{equation}\label{7a}
    \rho_{st}(\theta)=\dfrac{B(0)}{B(\theta)}\dfrac{f(\theta)}{f(0)}\rho_{st}(0)-2J\dfrac{f(\theta)}{B(\theta)}\int_{0}^{\theta}\dfrac{d\phi}{f(\phi)}.
\end{equation}
It is now easy to evaluate $J$ from the above equation. To this end, we use the periodic boundary condition $ \rho_{st}(0)=\rho_{st}(2\pi)$ in Eq.~\eqref{7a} to get
\begin{equation}\label{8a}
\rho_{st}(2\pi)=\dfrac{B(0)}{B(2\pi)}\dfrac{f(2\pi)}{f(0)}\rho_{st}(0)-2J\dfrac{f(2\pi)}{B(2\pi)}\int_{0}^{2\pi}\dfrac{d\phi}{f(\phi)},
\end{equation}
from which one trivially arrives at the following expression of the steady state current
\begin{equation}\label{9a}
    J=\dfrac{\left[\dfrac{B(0)}{f(0)}-\dfrac{B(2\pi)}{f(2\pi)}\right]\rho_{st}(0)}{2\int_{0}^{2\pi}\dfrac{d\phi}{f(\phi)}}.
\end{equation}
Substituting the current from Eq.~\eqref{9a} into Eq.~\eqref{7a}, one finally obtains
\begin{equation}\label{10a}
    \rho_{st}(\theta)=\dfrac{\dfrac{B(0)}{f(0)}\int_{\theta}^{2\pi}\dfrac{d\phi}{f(\phi)}+\dfrac{B(2\pi)}{f(2\pi)}\int_{0}^{\theta}\dfrac{d\phi}{f(\phi)}}{\dfrac{B(\theta)}{f(\theta)}\int_{0}^{2\pi}\dfrac{d\phi}{f(\phi)}}\rho_{st}(0),
\end{equation}
which is a general expression for the stationary state. Inspecting Eq.~\eqref{Eq.10}, one now easily identifies 
 $A(\theta)=Kr\sin \theta$ \& $B(\theta)=-D/2$. Finally, substituting all these in Eq.~\eqref{4a} and  
Eq.~\eqref{10a} we find the desired form of the stationary state
\begin{align}
       \rho_{st}(\theta)=\dfrac{\exp\left(\dfrac{Kr}{D}\cos\theta\right)}{2\pi I_0\left(\dfrac{Kr}{D}\right)},
\end{align}
which was announced in the main text.

%\bibliographystyle{apsrev4-1} % Tell bibtex which bibliography style to use
%\bibliography{author.bib}
%
\end{document}